\documentclass[aps,prl,10pt,twocolumn,superscriptaddress]{revtex4-1}
\usepackage{graphicx}
\usepackage{epsfig}
\usepackage{amsmath}
\usepackage{amssymb}
\usepackage{amsfonts}
\usepackage{mathrsfs}
\usepackage{theorem}
\usepackage{bm}
\usepackage{url}
\usepackage[T1]{fontenc}
\usepackage{csquotes}
\usepackage{natbib}
\usepackage{float}
\usepackage{appendix}
\usepackage{soul}
\usepackage[colorlinks=true,citecolor=blue,linkcolor=blue]{hyperref}
\MakeOuterQuote{"}

\begin{document}


\title{Beating the channel capacity limit for superdense coding with entangled ququarts}


\author{Xiao-Min Hu}
\affiliation{CAS Key Laboratory of Quantum Information, University of Science and Technology of China, Hefei, 230026, People's Republic of China}
\affiliation{Synergetic Innovation Center of Quantum Information and Quantum Physics, University of Science and Technology of China, Hefei, 230026, People's Republic of China}
\affiliation{These authors contribute equally to this work}

\author{Yu Guo}
\affiliation{CAS Key Laboratory of Quantum Information, University of Science and Technology of China, Hefei, 230026, People's Republic of China}
\affiliation{Synergetic Innovation Center of Quantum Information and Quantum Physics, University of Science and Technology of China, Hefei, 230026, People's Republic of China}
\affiliation{These authors contribute equally to this work}

\author{Bi-Heng Liu}
\email{bhliu@ustc.edu.cn}
\affiliation{CAS Key Laboratory of Quantum Information, University of Science and Technology of China, Hefei, 230026, People's Republic of China}
\affiliation{Synergetic Innovation Center of Quantum Information and Quantum Physics, University of Science and Technology of China, Hefei, 230026, People's Republic of China}

\author{Yun-Feng Huang}
\affiliation{CAS Key Laboratory of Quantum Information, University of Science and Technology of China, Hefei, 230026, People's Republic of China}
\affiliation{Synergetic Innovation Center of Quantum Information and Quantum Physics, University of Science and Technology of China, Hefei, 230026, People's Republic of China}

\author{Chuan-Feng Li}
\email{cfli@ustc.edu.cn}
\affiliation{CAS Key Laboratory of Quantum Information, University of Science and Technology of China, Hefei, 230026, People's Republic of China}
\affiliation{Synergetic Innovation Center of Quantum Information and Quantum Physics, University of Science and Technology of China, Hefei, 230026, People's Republic of China}

\author{Guang-Can Guo}
\affiliation{CAS Key Laboratory of Quantum Information, University of Science and Technology of China, Hefei, 230026, People's Republic of China}
\affiliation{Synergetic Innovation Center of Quantum Information and Quantum Physics, University of Science and Technology of China, Hefei, 230026, People's Republic of China}

\date{\today}

\begin{abstract}
Quantum superdense coding protocols enhance channel capacity by using shared quantum entanglement between two users. The channel capacity can be as high as $2$ when one uses entangled qubits. However, this limit can be surpassed by using high-dimensional entanglement. Here, we report an experiment that exceeds the limit using high-quality entangled ququarts with fidelities up to $0.98$, demonstrating a channel capacity of $2.09\pm0.01$. The measured channel capacity is also higher than that obtained when transmitting only one ququart. We use the setup to transmit a five-color image with a fidelity of $0.952$. Our experiment shows the great advantage of high-dimensional entanglement and will stimulate research on high-dimensional quantum information processes.
\end{abstract}

\maketitle
\section*{INTRODUCTION}
Quantum entanglement \cite{Horodecki} reveals the power of the quantum world. An important application of quantum entanglement is quantum superdense coding (SDC) \cite{Bennett}, relying on the shared quantum entanglement between Alice (sender) and Bob (receiver). Alice encodes 2 classical bits of information on her qubit and then sends the qubit to Bob; Bob performs a complete Bell-state measurement on the two qubits and decodes the 2 classical bits. During the process, Alice only sends 1 qubit to Bob, while Bob receives 2 bits of classical information. This protocol has been realized in many systems \cite{Mattle,Li,Fang,Schaetz}. Considering the photon as the natural flying qubit, the performance of SDC in a linear optical system attracts much attention. However, a complete Bell-state measurement with linear optics is forbidden \cite{Vaidman,Lutkenhaus}, and the theoretical channel capacity limit of 2 is hard to reach. Previous efforts were made to reach this limit \cite{Schuck,Barreiro,Williams} even in the presence of noise \cite{Liu}.

Compared with two-qubit entanglement, high-dimensional entanglement is more powerful and increases the channel capacity \cite{Liuxs,Grudka}. Recently, high-dimensional entanglement has been demonstrated in optical systems \cite{Dada,Schaeff,Krenn,Hu,Kues,Krenn18}. Based on a photon's orbital angular momentum, Zeilinger's group also demonstrated the preparation of a complete four-dimensional Bell-basis \cite{Wang} and quantum gates on single high-dimensional photons \cite{Babazadeh}. These techniques pave the way to high-dimensional quantum communication. However, the fidelity of the high-dimensional Bell-state in the previous works is slightly lower, such as $0.78\pm0.03$ \cite{Wang}, and thus, it is hard to carry out the SDC protocol. On the other hand, we have used beam displacers to build very stable Mach-Zehnder interferometers \cite{Hu,OBrien,Liu2} and obtained a high-quality three-dimensional path entanglement with a fidelity of $0.975\pm0.001$ \cite{Hu}. In this article, we report on an experimental demonstration of SDC using path-polarization-encoded entanglement. By encoding the spatial mode and polarization mode of the photon, the system is more compact than the previous one \cite{Hu}, and the fidelity of the four-dimensional entangled state is up to $0.980\pm0.001$. The measured channel capacity is $2.09\pm0.01$, which exceeds the limit of 2 from the qubit system for the first time \cite{Bennett}. Moreover, the channel capacity is also higher than the limit of transmitting one ququart \cite{Milione}. We use the setup to transmit a real image and observe a fidelity of $0.952$. This makes the efficient SDC using high-dimensional entanglement feasible in the future.

\section*{RESULTS}
\begin{figure*}[tbph]
\begin{center}
\includegraphics [width= 1.6 \columnwidth]{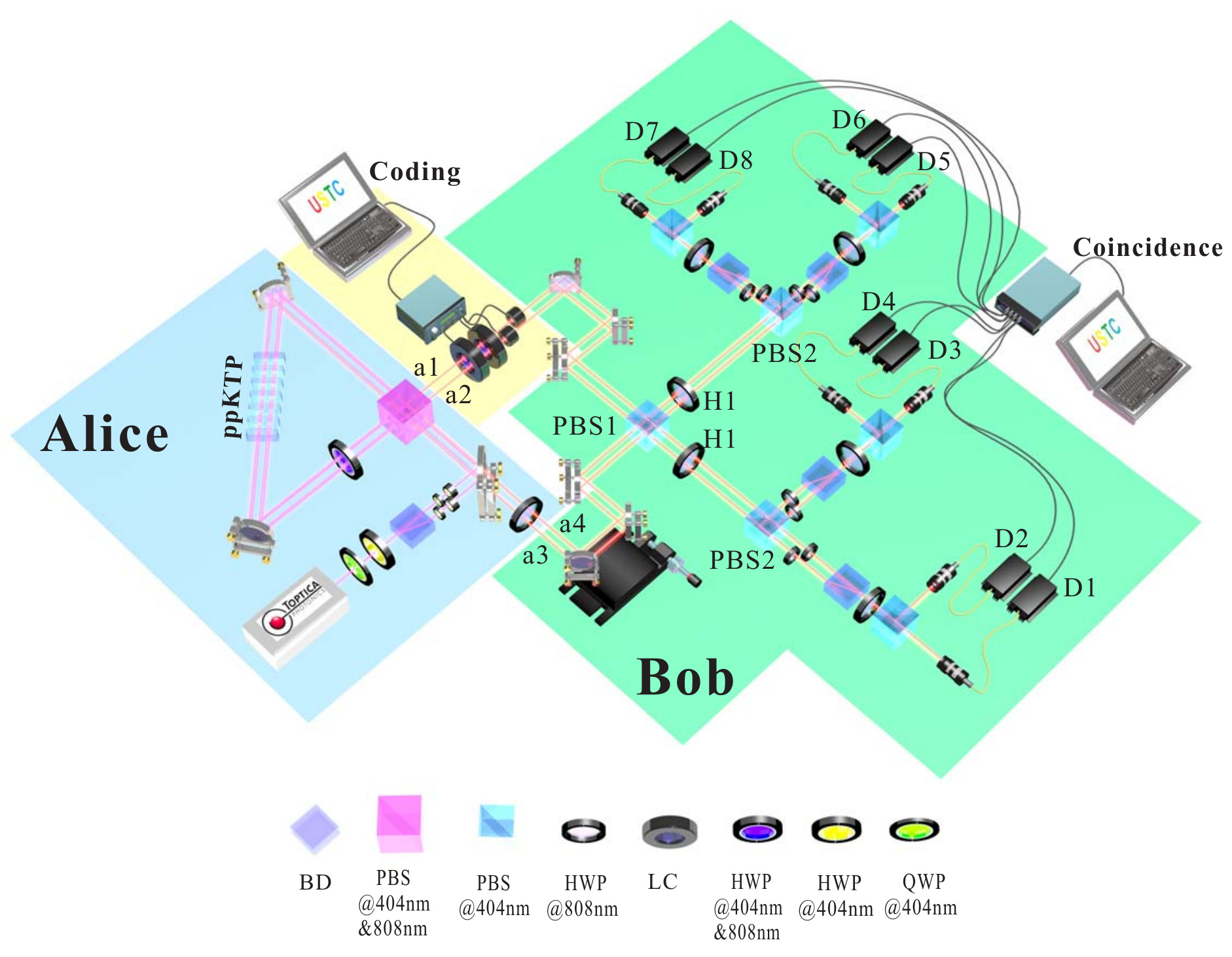}
\caption{{\bf Experimental setup.} A CW violet laser (power is $4$ mW, wavelength is $404$ nm) is focused by two lenses and the waist radius is approximate 0.25 mm. Then the light beam is separated into two paths by a beam displacer. These two beams are injected into a Sagnac interferometer to pump a type-II cut periodically poled potassium titanyl phosphate (ppKTP) crystal ($1$ mm $\times$ $7$ mm $\times$ $10$ mm), and generate two photon polarization entanglement $(|H\rangle|H\rangle+|V\rangle|V\rangle)/\sqrt{2}$ in each path \cite{Fedrizzi}. The ppKTP is temperature controlled by a home-made temperature controller, and the temperature stability is $0.001$ K to ensure the phase stability between the two paths. Then, we encode horizontally polarized ($H$) photons in path a1 (a3) as $|0\rangle$, vertically polarized ($V$) photons in path a1 (a3) as $|1\rangle$, $H$ photons in path a2 (a4) as $|2\rangle$, $V$ photons in path a2 (a4) as $|3\rangle$, and carefully adjust the relative phase between the two paths, the state is prepared in a four-dimensional maximally entangled two-photon state $\Psi_{11}=(|00\rangle+|11\rangle+|22\rangle+|33\rangle)/2$. Finally, Alice encodes her information using four computer-controlled liquid crystal variable retarders and sends her photon to Bob. Bob performs a measurement on the two photons and decodes the information that Alice encoded, details can be found in the Supplementary Materials. BD-beam displacer, PBS-polarizing beam splitter, HWP-half wave plate, LC-liquid crystal variable retarder.}
\end{center}\label{fig.1}
\end{figure*}

To generate high-quality, high-dimensional entangled states, we use a path-polarization hybrid system as shown in Figure 1. We encode horizontally polarized ($H$) photons in path a1 (a3) as $|0\rangle$, vertically polarized ($V$) photons in path a1 (a3) as $|1\rangle$, $H$ photons in path a2 (a4) as $|2\rangle$, $V$ photons in path a2 (a4) as $|3\rangle$ (here $|0\rangle$, $|1\rangle$, $|2\rangle$ and $|3\rangle$ represent high-dimensional bases, not Fock states). Thus, the two-photon state is $\Psi_{11}=(|00\rangle+|11\rangle+|22\rangle+|33\rangle)/2$. Then, Alice uses four computer-controlled liquid crystal variable retarders (LCs, which introduces birefringence between fast axis and slow axis depending on the voltage applied on it) to perform local operations on her own photon and sends it to Bob for Bell-state measurement (Details can be found in Materials and Methods). Although, it is impossible to perform a complete Bell-state measurement on a four-dimensional system with linear optics \cite{Calsamiglia}, we can separate the 16 Bell-states into seven classes and select only one state in each class for SDC with the help of photon-number resolving detectors \cite{Hill,Wei}. In our system, we choose five states for SDC
\begin{equation}\label{1}
\begin{split}
\begin{aligned}
\Psi_{11}=\frac{1}{2}(|00\rangle+|11\rangle+|22\rangle+|33\rangle),  \\
\Psi_{12}=\frac{1}{2}(|00\rangle-|11\rangle+|22\rangle-|33\rangle),  \\
\Psi_{13}=\frac{1}{2}(|00\rangle+|11\rangle-|22\rangle-|33\rangle),  \\
\Psi_{14}=\frac{1}{2}(|00\rangle-|11\rangle-|22\rangle+|33\rangle),  \\
\Psi_{23}=\frac{1}{2}(|01\rangle+|10\rangle-|23\rangle-|32\rangle). \\
\end{aligned}
\end{split}
\end{equation}

The channel capacity using five-state SDC is $\log_2{5}\simeq2.32$, which is higher than the limit of $2$ set by using entangled qubits and is also higher than the limit of transmitting one qudit. By carefully checking the coincidental events among those single photon detectors, Bob can determine which state Alice prepared, details see the Supplementary Materials. For example, coincidence between D1 and D5 means the state Alice prepared is $\Psi_{11}$.

\begin{figure}[tbph]
\begin{center}
\includegraphics [width= 1 \columnwidth]{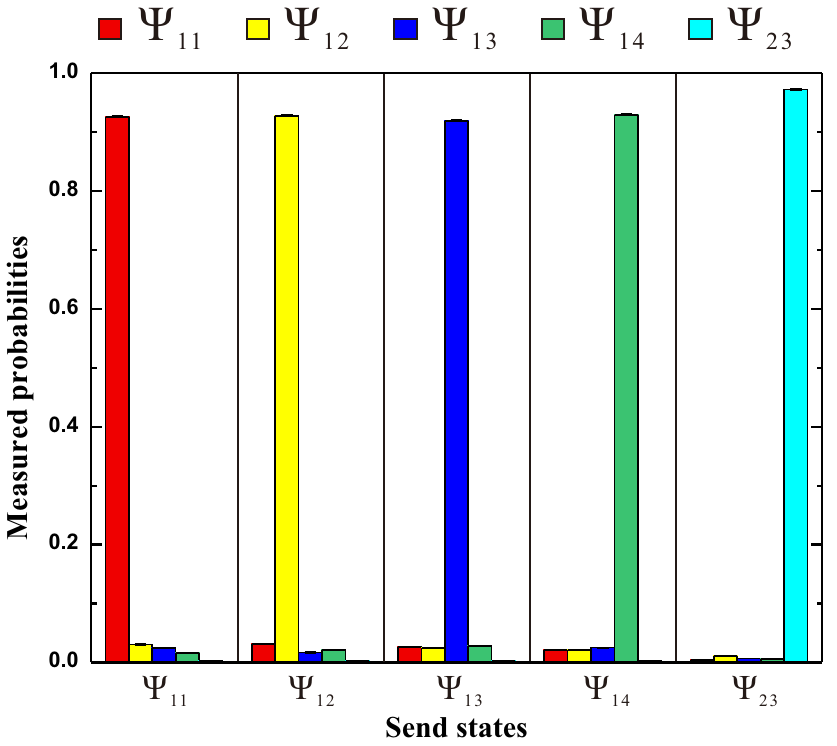}
\caption{{\bf Measured probabilities.} Alice sends the five four-dimensional Bell-states to Bob, and Bob performs a measurement on the two photons and obtains the probabilities for each state. In our experiment, the photon-pair count rate is 1000 /s, and the integration time is 20 s for each input state. Error bars are due to the statistical error.}
\end{center}\label{fig2}
\end{figure}


First, we test the state preparation and Bell-state measurement. Alice sends each state to Bob, and Bob performs a measurement and checks the success probabilities. The results are shown in Figure 2. For states $\Psi_{11}$, $\Psi_{12}$, $\Psi_{13}$ and $\Psi_{14}$, the average success probability is $0.926\pm0.002$, and for state $\Psi_{23}$, the success probability is $0.972\pm0.002$. This is because the first four states suffer from imperfect Hong-Ou-Mandel interference, while the last is just a polarization projection. In our experiment, the measured visibility of Hong-Ou-Mandel interference is $0.962\pm0.002$ when using a $10$-mm-long ppKTP crystal. Hence, the channel capacity in our system is measured to be $2.09\pm0.01$, which is slightly lower than the theoretical prediction of $2.32$ but still exceeds the limit when using the entangled qubit system. When one transmits one ququart, the channel capacity is $\log_2{4}=2$. Our result also exceeds this limit. This means that our experiment proves that high-dimensional SDC can increase the channel capacity; thus, it is possible to perform quantum communication via efficient SDC even in a high-dimensional system.

\begin{figure}[tbph]
\begin{center}
\includegraphics [width= 0.9 \columnwidth]{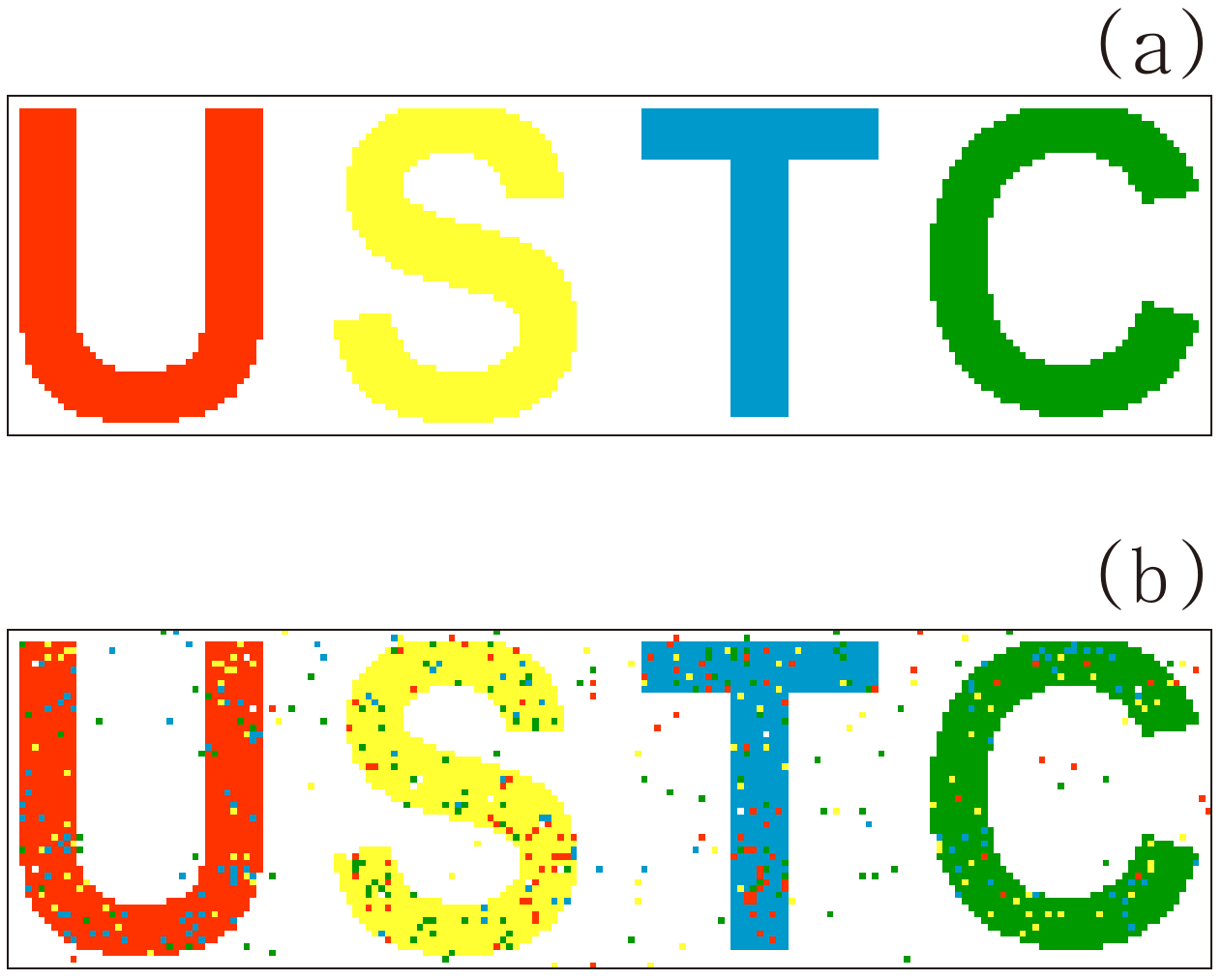}
\caption{{\bf Transmit a real five-color image using encoded information.} Here red spots are encoded to $\Psi_{11}$, yellow spots are encoded to $\Psi_{12}$, blue spots are encoded to $\Psi_{13}$, green spots are encoded to $\Psi_{14}$, and white spots are encoded to $\Psi_{23}$. (a) The original 5-colour $53\times188$ pixel image. (b) The image received using superdense coding. The calculated fidelity is 0.952.}
\end{center}\label{Fig3}
\end{figure}

Then, we use the encoded information to transmit a real image as shown in Figure 3. Alice encodes a five-color image on her own photon (each color corresponds to a four-dimensional Bell-state, and she can transform the initial state to the others by four-computer-controlled LCs. Alice's encoding process can be found in Materials and Methods) and then transmits the encoded photon to Bob. Bob receives the photon and performs a measurement on the two photons to determine which state Alice encoded and then decodes the information. In our experiment, Alice performs single photon operation on her own photon by four computer-controlled LCs and sends it to Bob, this process will take $0.5$ second; then, Bob starts to measure the two-photon state after $0.5$ second until he obtains a two-fold coincident event within $1$ second. These process is repeated with a repetition rate of $0.5$ Hz and the SDC data rate is $0.5$ Hz. Theoretically, our setup can distinguish all five of the four-dimensional Bell-states with a single-shot measurement when the detection efficiency is $100\%$. However, the overall detection efficiency is $0.109\pm0.002$ in our experiment due to the loss of detectors (approximate 50\% detection efficiency at 808 nm) and coupling efficiency (approximate 22\% including the losses of optical elements), thus Bob needs more time to ensure that he has completed the measurement.

\section{DISCUSSION}

SDC with high dimensional entanglement is promising in the future. Note that since the photon pairs are generated from ppKTP, the overall detection efficiency can be as high as $78.6\%$ when using careful alignment and superconductor detectors \cite{Giustina}. The computer-controlled LC operation rate is approximately $30$ Hz; this part can be replaced by a fast electro-optical modulator with an operation rate of $1$ GHz. Together with a pulsed laser source with a repetition rate of tens of MHz or even faster, the SDC data rate can be as high as $1$ MHz. The path-polarization hyperentanglement can also be distributed to distant users by a multi-core fibre \cite{Ding}, and our SDC with high-dimensional entanglement is feasible in the future for fast, distant quantum communication. Further-more, path-polarization hyperentanglement is possible to connect to integrated photonic chips \cite{Politi,Wang2018}.

In conclusion, we demonstrate the SDC protocol using four-dimensional entanglement for the first time and achieve a channel capacity of $2.09\pm0.01$, which exceeds the limit of SDC using two-dimensional entanglement and the limit from transmitting one ququart. This experiment will stimulate the research of applications based on high-dimensional entanglement.

\section{MATERIALS AND METHODS}

{\bf Alice's encoding process.}
The initial state shared by Alice and Bob is $\Psi_{11}=\frac{1}{2}(|00\rangle+|11\rangle+|22\rangle+|33\rangle)$, and Alice performs single-photon operations on her own photon and realizes the state transformation from $\Psi_{11}$ to the other 4 Bell-states. These operations are realized by four computer-controlled LCs as shown in Figure 4. Then, Alice can easily control the state by applying different voltages on the LCs.
If all the LCs are set to introduce 0 phase, the two-photon state is
\begin{equation}
\Psi_{11}=\frac{1}{2}(|00\rangle+|11\rangle+|22\rangle+|33\rangle).
\end{equation}
If LC2 is set to introduce $\pi$ phase, while the others remain at 0, then the two-photon state is
\begin{equation}
\Psi_{12}=\frac{1}{2}(|00\rangle-|11\rangle+|22\rangle-|33\rangle).
\end{equation}
If LC2, LC3 and LC4 are set to introduce $\pi$ phase, while LC1 remains at 0, then the two-photon state is
\begin{equation}
\Psi_{13}=\frac{1}{2}(|00\rangle+|11\rangle-|22\rangle-|33\rangle).
\end{equation}
If LC3 and LC4 are set to introduce $\pi$ phase, while LC1 and LC2 remain at 0, then the two-photon state is
\begin{equation}
\Psi_{14}=\frac{1}{2}(|00\rangle-|11\rangle-|22\rangle+|33\rangle).
\end{equation}
If all the LCs are set to introduce $\pi$ phase, then the two-photon state is
\begin{equation}
\Psi_{23}=\frac{1}{2}(|01\rangle+|10\rangle-|23\rangle-|32\rangle).
\end{equation}
\begin{figure}[tbph]
\begin{center}
\includegraphics [width= 0.9 \columnwidth]{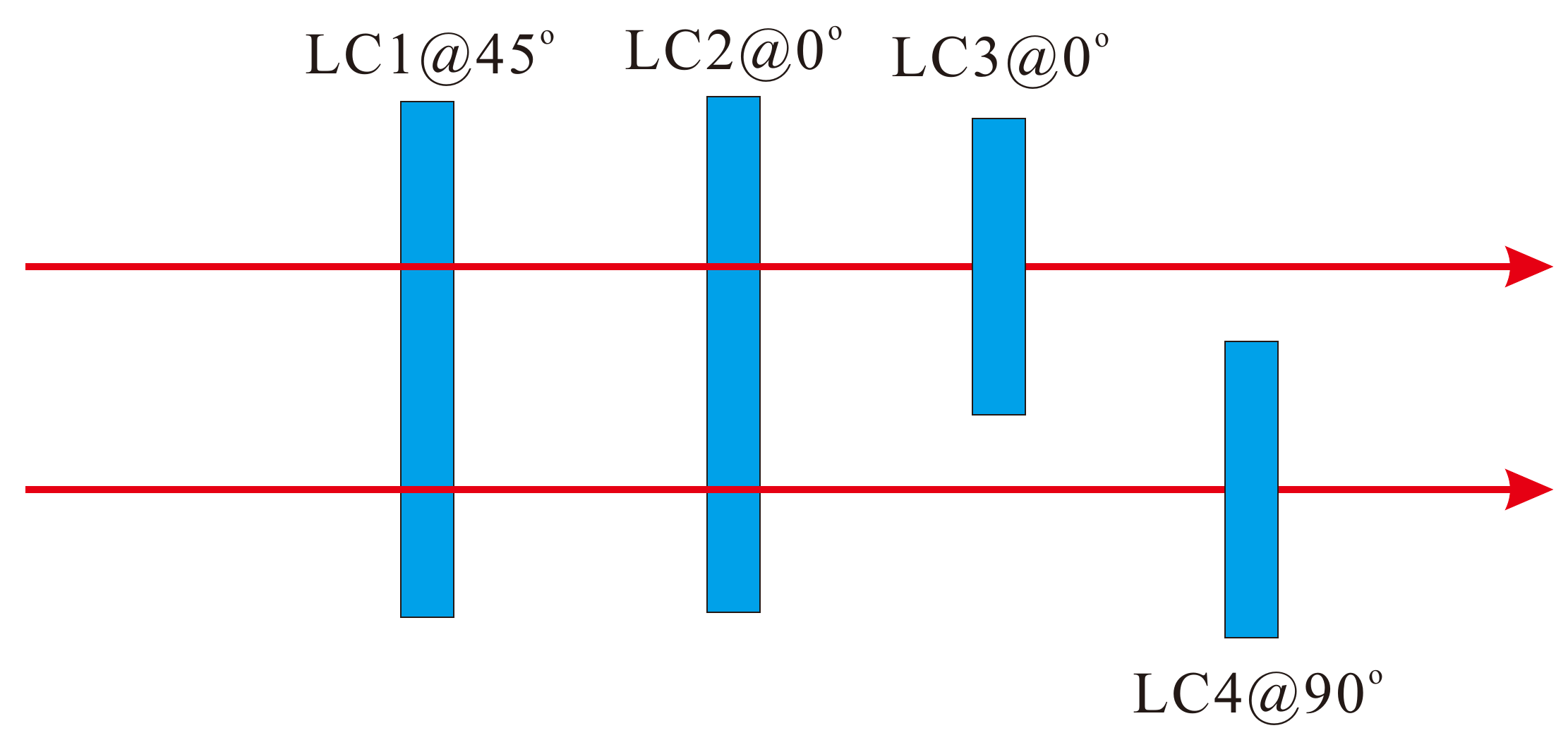}
\caption{{\bf Realization of a single-photon operation on Alice's side.} The optical axes of the liquid crystal variable retarders (LCs) are set at different angles as shown in the figure. By applying different voltages, the LCs will introduce different phases between the fast axis and the slow axis.}
\end{center}\label{fig4}
\end{figure}

{\bf Bell-state measurement.}
A complete four-dimensional Bell-state measurement is impossible with linear elements \cite{Calsamiglia}. However, one can separate the 16 four-dimensional Bell-states into seven classes and select only one state from each class for SDC \cite{Hill,Wei}. In our experiment, we only choose five states given in Eq. \ref{1}. Then Bob can determine which state Alice prepared by the measurement setup shown in Fig. 1. When the two photons inject into PBS1, they are separated into two classes by photon's polarization. For state $\Psi_{11}$, $\Psi_{12}$, $\Psi_{13}$, $\Psi_{14}$, one photon is reflected and the other one is transmitted, while for state $\Psi_{23}$, both photons are reflected or transmitted. After half wave plate H1 (set at $22.5^\circ$) and PBS2, state $\Psi_{11}$, $\Psi_{12}$, $\Psi_{13}$, $\Psi_{14}$ are separated into two classes by the relative phase between $|00\rangle$ and $|11\rangle$, for example, if the relative phase is 0 ($\Psi_{11}$, $\Psi_{13}$), then the two photons will both transmit or reflect from PBS2. Here, this separation depends on the two-photon Hong-Ou-Mandel interference \cite{Liu}. Finally, the last part of measurement setup (consists of HWPs, BDs and PBSs) will separate $\Psi_{11}$ and $\Psi_{13}$ ($\Psi_{12}$ and $\Psi_{14}$) to two classes by the relative phase between $|00\rangle$ and $|22\rangle$. By carefully checking the coincidental events among those single photon detectors, Bob can determine which state Alice prepared. For example, coincidence between D1 and D5 means the state Alice prepared is $\Psi_{11}$. Thus, Bob can determine which state Alice prepared and decode the information that Alice encoded.

{\bf Channel capacity.}
The channel capacity \cite{Barreiro,Williams} is the maximal mutual information $H(x;y)$. In our experiment, the mutual information can be calculated as
\begin{equation}
H(x;y)=\sum\limits_{y=0}^{4}\sum\limits_{x=0}^{4}p(x)p(y|x)log(\frac{p(y|x)}{\sum\limits_{x=0}^{4}p(x)p(y|x)}),
\end{equation}
where $p(x)=\{p(\Psi_{11}),p(\Psi_{12}),p(\Psi_{13}),p(\Psi_{14}),p(\Psi_{23})\}$. The channel capacity in our experiment is $2.09\pm0.01$, which is slightly smaller than the limit of $2.32$ owing to the imperfect four-dimensional Bell-state measurement.




\section*{ACKNOWLEDGMENTS}
{\bf Funding:} This work was supported by the National Key Research and Development Program of China (No.\ 2017YFA0304100), NNSFC (Nos.\ 11374288, 11774335, 61327901, 11474268, 11325419, 11504253), the Key Research Program of Frontier Sciences, CAS (No.\ QYZDY-SSW-SLH003), the Fundamental Research Funds for the Central Universities, and Anhui Initiative in Quantum Information Technologies (Nos.\ AHY020100, AHY060300). {\bf Author contributions:} The experiment was performed by X-M.H., B-H.L., Y.G.. The data analysis was performed by X-M.H., B-H.L. and C-F.L. with input from all authors. The paper was written by X-M.H., B-H.L., and C-F.L. All authors discussed the experimental results. {\bf Competing interests:} The authors declare that they have no competing interests. {\bf Data and materials availability:} All data needed to evaluate the conclusions in the paper are present in the paper and/or the Supplementary Materials. Additional data related to this paper may be requested from the authors.

\bigskip
\bigskip
\bigskip
\appendix
\onecolumngrid
\section{Appendix}
\setcounter{figure}{0}
\renewcommand{\thefigure}{A\arabic{figure}}
\setcounter{equation}{0}
\renewcommand{\theequation}{A\arabic{equation}}
{\em Two-photon Four-dimensional Bell-basis.---} For a four-dimensional bipartite system, the complete Bell-basis can be written as \cite{Wang}
\begin{equation}\
\begin{split}
\begin{aligned}
\Psi_{11}=\frac{1}{2}(|00\rangle+|11\rangle+|22\rangle+|33\rangle),  \\
\Psi_{12}=\frac{1}{2}(|00\rangle-|11\rangle+|22\rangle-|33\rangle),  \\
\Psi_{13}=\frac{1}{2}(|00\rangle+|11\rangle-|22\rangle-|33\rangle),  \\
\Psi_{14}=\frac{1}{2}(|00\rangle-|11\rangle-|22\rangle+|33\rangle),  \\
\Psi_{21}=\frac{1}{2}(|01\rangle+|10\rangle+|23\rangle+|32\rangle),  \\
\Psi_{22}=\frac{1}{2}(|01\rangle-|10\rangle+|23\rangle-|32\rangle),  \\
\Psi_{23}=\frac{1}{2}(|01\rangle+|10\rangle-|23\rangle-|32\rangle),  \\
\Psi_{24}=\frac{1}{2}(|01\rangle-|10\rangle-|23\rangle+|32\rangle),  \\
\Psi_{31}=\frac{1}{2}(|02\rangle+|13\rangle+|20\rangle+|31\rangle),  \\
\Psi_{32}=\frac{1}{2}(|02\rangle-|13\rangle+|20\rangle-|31\rangle),  \\
\Psi_{33}=\frac{1}{2}(|02\rangle+|13\rangle-|20\rangle-|31\rangle),  \\
\Psi_{34}=\frac{1}{2}(|02\rangle-|13\rangle-|20\rangle+|31\rangle),  \\
\Psi_{41}=\frac{1}{2}(|03\rangle+|12\rangle+|21\rangle+|30\rangle),  \\
\Psi_{42}=\frac{1}{2}(|03\rangle-|12\rangle+|21\rangle-|30\rangle),  \\
\Psi_{43}=\frac{1}{2}(|03\rangle+|12\rangle-|21\rangle-|30\rangle),  \\
\Psi_{44}=\frac{1}{2}(|03\rangle-|12\rangle-|21\rangle+|30\rangle). \\
\end{aligned}
\end{split}
\end{equation}

{\em Bell-state measurement.---} In the quantum superdense coding protocol, one need to distinguish all the 16 Bell-basis in a single-shot measurement. However, this is impossible with linear optics \cite{Calsamiglia}. A reliable way is to select some of the Bell-basis and determine the basis by a single-shot measurement. Here, we show how to select the special bases and how to distinguish them with linear optics \cite{Hill,Wei}.

The experimental setup is shown in Fig. A1.

\begin{figure}
\centering
\includegraphics[width=0.6\textwidth]{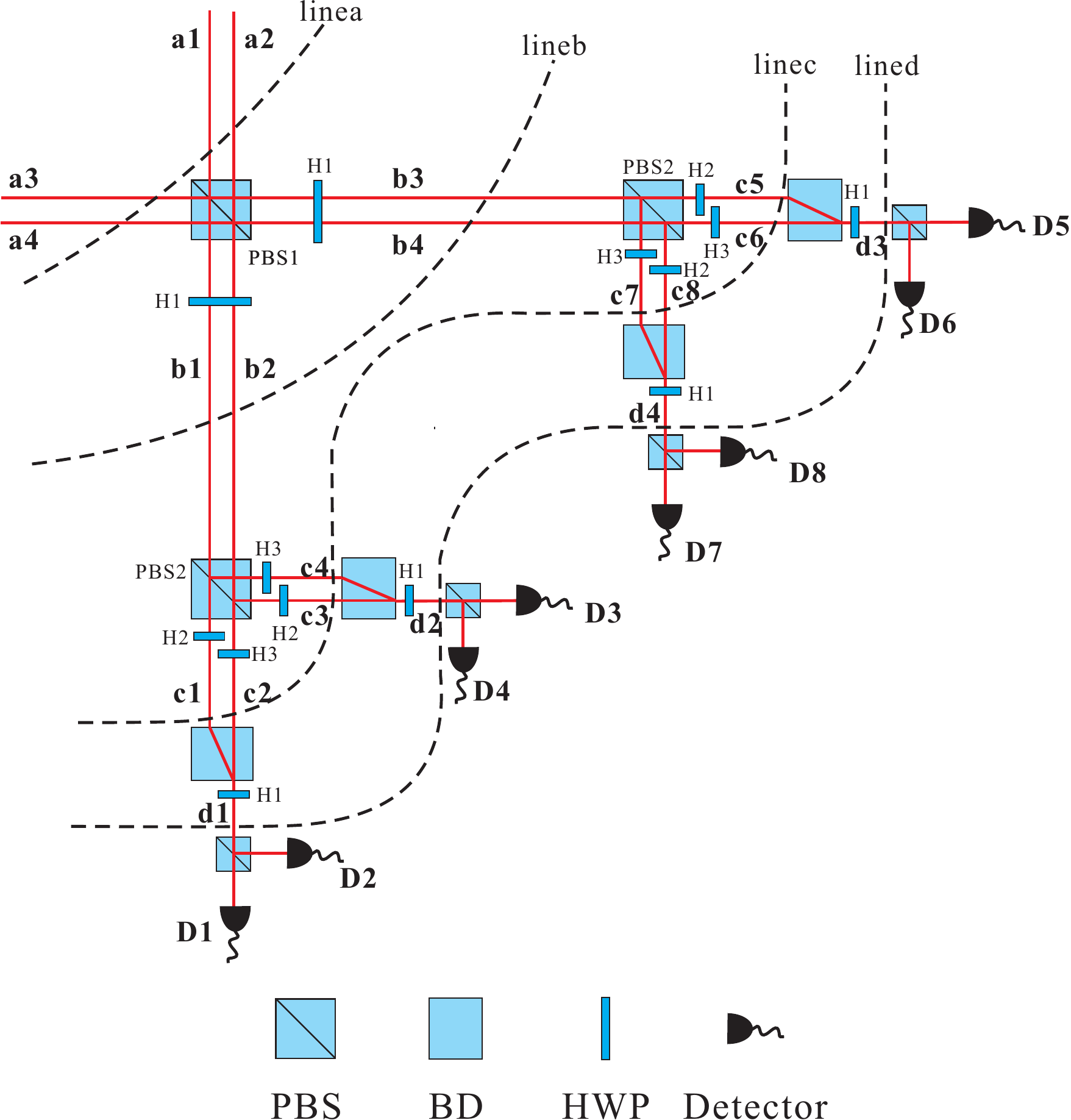}
\caption{{\bf Measurement on four-dimensional two-photon Bell-state.} We divide the whole process to five part by four dashed lines. We encode the horizontally polarized ($H$) photon in path a1 (a3) as $|0\rangle$, the vertically polarized ($V$) photon in path a1 (a3) as $|1\rangle$, and the $H$ photon in path a2 (a4) as $|2\rangle$, the $V$ photon in path a2 (a4) as $|3\rangle$. Half wave plates (HWPs) are set at special degrees to rotate the polarizations of photons or compensate the optical path difference. Here, H1 is set at $22.5^\circ$, H2 is set at $0^\circ$ and H3 is set at $45^\circ$. PBS-polarizing beam splitter, BD-beam displacer, HWP-half wave plate.}
\end{figure}

For each input four-dimensional two-photon state, the measurement result can be described as following.

1. If the input state is $\Psi_{11}=\frac{1}{2}(|00\rangle+|11\rangle+|22\rangle+|33\rangle)$, then we can rewrite the state before each line
\begin{equation}
\begin{split}
\begin{aligned}
&\Psi_{11}=\frac{1}{2}(|00\rangle+|11\rangle+|22\rangle+|33\rangle)  \\
&\xrightarrow{line a}\frac{1}{2}(|H_{a1}H_{a3}\rangle+|V_{a1}V_{a3}\rangle+|H_{a2}H_{a4}\rangle+|V_{a2}V_{a4}\rangle)  \\
&\xrightarrow{line b}\frac{1}{2}(\frac{1}{2}(|H_{b1}+V_{b1}\rangle\ |H_{b3}+V_{b3}\rangle)+\frac{1}{2}(|H_{b1}-V_{b1}\rangle\ |H_{b3}-V_{b3}\rangle)) \\
&+\frac{1}{2}(|H_{b2}+V_{b2}\rangle\ |H_{b4}+V_{b4}\rangle)+\frac{1}{2}(|H_{b2}-V_{b2}\rangle\ |H_{b4}-V_{b4}\rangle)   \\
&=\frac{1}{4}(|H_{b1}H_{b3}\rangle+|H_{b1}V_{b3}\rangle+|V_{b1}H_{b3}\rangle+|V_{b1}V_{b3}\rangle+|H_{b1}H_{b3}\rangle-|H_{b1}V_{b3}\rangle-|V_{b1}H_{b3}\rangle+|V_{b1}V_{b3}\rangle  \\
&+|H_{b2}H_{b4}\rangle+|H_{b2}V_{b4}\rangle+|V_{b2}H_{b4}\rangle+|V_{b2}V_{b4}\rangle+|H_{b2}H_{b4}\rangle-|H_{b2}V_{b4}\rangle-|V_{b2}H_{b4}\rangle+|V_{b2}V_{b4}\rangle)  \\
&=\frac{1}{2}(|H_{b1}H_{b3}\rangle+|V_{b1}V_{b3}\rangle+|H_{b2}H_{b4}\rangle+|V_{b2}V_{b4}\rangle) \\
&\xrightarrow{line c}\frac{1}{2}(|H_{c1}H_{c5}\rangle+|V_{c2}V_{c6}\rangle+|H_{c4}H_{c7}\rangle+|V_{c3}H_{c8}\rangle) \\
&\xrightarrow{line d}\frac{1}{2}(|H_{d1}H_{d3}\rangle+|V_{d1}V_{d3}\rangle+|H_{d2}H_{d4}\rangle+|V_{d2}H_{d4}\rangle) \\
&\xrightarrow{} C_{D1,D5}+C_{D2,D6}+C_{D3,D7}+C_{D8,D4},
\end{aligned}
\end{split}
\end{equation}
here, $H_{a1}$ means a horizontally polarized photon at mode $a1$, $C_{D1,D5}$ means two folds coincidence between detector $D1$ and $D5$. One can find that if the input state is $\Psi_{11}$, then there have only four possible coincidental events, such as, $C_{D1,D5}$, $C_{D2,D6}$, $C_{D3,D7}$ and $C_{D8,D4}$.

2. If the input state is $\Psi_{12}=\frac{1}{2}(|00\rangle-|11\rangle+|22\rangle-|33\rangle)$, then we can rewrite the state before each line
\begin{equation}
\begin{split}
\begin{aligned}
&\Psi_{12}=\frac{1}{2}(|00\rangle-|11\rangle+|22\rangle-|33\rangle)  \\
&\xrightarrow{line a}\frac{1}{2}(|H_{a1}H_{a3}\rangle-|V_{a1}V_{a3}\rangle+|H_{a2}H_{a4}\rangle-|V_{a2}V_{a4}\rangle)  \\
&\xrightarrow{line b}\frac{1}{2}(\frac{1}{2}(|H_{b1}+V_{b1}\rangle\ |H_{b3}+V_{b3}\rangle)-\frac{1}{2}(|H_{b1}-V_{b1}\rangle\ |H_{b3}-V_{b3}\rangle)) \\
&+\frac{1}{2}(|H_{b2}+V_{b2}\rangle\ |H_{b4}+V_{b4}\rangle)-\frac{1}{2}(|H_{b2}-V_{b2}\rangle\ |H_{b4}-V_{b4}\rangle)   \\
&=\frac{1}{4}(|H_{b1}H_{b3}\rangle+|H_{b1}V_{b3}\rangle+|V_{b1}H_{b3}\rangle+|V_{b1}V_{b3}\rangle-|H_{b1}H_{b3}\rangle+|H_{b1}V_{b3}\rangle+|V_{b1}H_{b3}\rangle-|V_{b1}V_{b3}\rangle  \\
&+|H_{b2}H_{b4}\rangle+|H_{b2}V_{b4}\rangle+|V_{b2}H_{b4}\rangle+|V_{b2}V_{b4}\rangle-|H_{b2}H_{b4}\rangle+|H_{b2}V_{b4}\rangle+|V_{b2}H_{b4}\rangle-|V_{b2}V_{b4}\rangle)  \\
&=\frac{1}{2}(|H_{b1}V_{b3}\rangle+|V_{b1}H_{b3}\rangle+|H_{b2}V_{b4}\rangle+|V_{b2}H_{b4}\rangle) \\
&\xrightarrow{line c}\frac{1}{2}(|H_{c1}H_{c7}\rangle+|V_{c2}V_{c8}\rangle+|H_{c4}H_{c5}\rangle+|V_{c3}V_{c6}\rangle) \\
&\xrightarrow{line d}\frac{1}{2}(|H_{d1}H_{d4}\rangle+|V_{d1}V_{d4}\rangle+|H_{d2}H_{d3}\rangle+|V_{d2}V_{d3}\rangle) \\
&\xrightarrow{} C_{D1,D7}+C_{D2,D8}+C_{D5,D3}+C_{D6,D4}.
\end{aligned}
\end{split}
\end{equation}

3. If the input state is $\Psi_{13}=\frac{1}{2}(|00\rangle+|11\rangle-|22\rangle-|33\rangle)$, then we can rewrite the state before each line
\begin{equation}
\begin{split}
\begin{aligned}
&\Psi_{13}=\frac{1}{2}(|00\rangle+|11\rangle-|22\rangle-|33\rangle)  \\
&\xrightarrow{line a}\frac{1}{2}(|H_{a1}H_{a3}\rangle+|V_{a1}V_{a3}\rangle-|H_{a2}H_{a4}\rangle-|V_{a2}V_{a4}\rangle)  \\
&\xrightarrow{line b}\frac{1}{2}(\frac{1}{2}(|H_{b1}+V_{b1}\rangle\ |H_{b3}+V_{b3}\rangle)+\frac{1}{2}(|H_{b1}-V_{b1}\rangle\ |H_{b3}-V_{b3}\rangle)) \\
&-\frac{1}{2}(|H_{b2}+V_{b2}\rangle\ |H_{b4}+V_{b4}\rangle)-\frac{1}{2}(|H_{b2}-V_{b2}\rangle\ |H_{b4}-V_{b4}\rangle)   \\
&=\frac{1}{4}(|H_{b1}H_{b3}\rangle+|H_{b1}V_{b3}\rangle+|V_{b1}H_{b3}\rangle+|V_{b1}V_{b3}\rangle+|H_{b1}H_{b3}\rangle-|H_{b1}V_{b3}\rangle-|V_{b1}H_{b3}\rangle+|V_{b1}V_{b3}\rangle  \\
&-|H_{b2}H_{b4}\rangle-|H_{b2}V_{b4}\rangle-|V_{b2}H_{b4}\rangle-|V_{b2}V_{b4}\rangle-|H_{b2}H_{b4}\rangle+|H_{b2}V_{b4}\rangle+|V_{b2}H_{b4}\rangle-|V_{b2}V_{b4}\rangle)  \\
&=\frac{1}{2}(|H_{b1}H_{b3}\rangle+|V_{b1}V_{b3}\rangle-|H_{b2}H_{b4}\rangle-|V_{b2}V_{b4}\rangle) \\
&\xrightarrow{line c}\frac{1}{2}(|H_{c1}H_{c5}\rangle-|V_{c2}V_{c6}\rangle+|H_{c4}H_{c7}\rangle-|V_{c3}V_{c8}\rangle) \\
&\xrightarrow{line d}\frac{1}{2}(|H_{d1}H_{d3}\rangle-|V_{d1}V_{d3}\rangle+|H_{d2}H_{d4}\rangle-|V_{d2}V_{d4}\rangle) \\
&\xrightarrow{} C_{D1,D6}+C_{D2,D5}+C_{D8,D4}+C_{D3,D7}.
\end{aligned}
\end{split}
\end{equation}

4. If the input state is $\Psi_{14}=\frac{1}{2}(|00\rangle-|11\rangle-|22\rangle+|33\rangle)$, then we can rewrite the state before each line
\begin{equation}
\begin{split}
\begin{aligned}
&\Psi_{14}=\frac{1}{2}(|00\rangle-|11\rangle-|22\rangle+|33\rangle)  \\
&\xrightarrow{line a}\frac{1}{2}(|H_{a1}H_{a3}\rangle-|V_{a1}V_{a3}\rangle-|H_{a2}H_{a4}\rangle+|V_{a2}V_{a4}\rangle)  \\
&\xrightarrow{line b}\frac{1}{2}(\frac{1}{2}(|H_{b1}+V_{b1}\rangle\ |H_{b3}+V_{b3}\rangle)-\frac{1}{2}(|H_{b1}-V_{b1}\rangle\ |H_{b3}-V_{b3}\rangle)) \\
&-\frac{1}{2}(|H_{b2}+V_{b2}\rangle\ |H_{b4}+V_{b4}\rangle)+\frac{1}{2}(|H_{b2}-V_{b2}\rangle\ |H_{b4}-V_{b4}\rangle)   \\
&=\frac{1}{4}(|H_{b1}H_{b3}\rangle+|H_{b1}V_{b3}\rangle+|V_{b1}H_{b3}\rangle+|V_{b1}V_{b3}\rangle-|H_{b1}H_{b3}\rangle+|H_{b1}V_{b3}\rangle+|V_{b1}H_{b3}\rangle-|V_{b1}V_{b3}\rangle  \\
&-|H_{b2}H_{b4}\rangle-|H_{b2}V_{b4}\rangle-|V_{b2}H_{b4}\rangle-|V_{b2}V_{b4}\rangle+|H_{b2}H_{b4}\rangle-|H_{b2}V_{b4}\rangle-|V_{b2}H_{b4}\rangle+|V_{b2}V_{b4}\rangle)  \\
&=\frac{1}{2}(|H_{b1}V_{b3}\rangle+|V_{b1}H_{b3}\rangle-|H_{b2}V_{b4}\rangle-|V_{b2}H_{b4}\rangle) \\
&\xrightarrow{line c}\frac{1}{2}(|H_{c1}H_{c7}\rangle+|V_{c2}V_{c8}\rangle-|H_{c4}H_{c5}\rangle-|V_{c3}V_{c6}\rangle) \\
&\xrightarrow{line d}\frac{1}{2}(|H_{d1}H_{D4}\rangle-|V_{d1}V_{d4}\rangle+|H_{d2}H_{d3}\rangle-|V_{d2}V_{d3}\rangle) \\
&\xrightarrow{} C_{D1,D8}+C_{D2,D7}+C_{D4,D5}+C_{D3,D6}.
\end{aligned}
\end{split}
\end{equation}

5. If the input state is $\Psi_{23}=\frac{1}{2}(|01\rangle+|10\rangle-|23\rangle-|32\rangle)$, then we can rewrite the state before each line
\begin{equation}
\begin{split}
\begin{aligned}
&\Psi_{23}=\frac{1}{2}(|01\rangle+|10\rangle-|23\rangle-|32\rangle)  \\
&\xrightarrow{line a}\frac{1}{2}(|H_{a1}V_{a3}\rangle+|V_{a1}H_{a3}\rangle-|H_{a2}V_{a4}\rangle-|V_{a2}H_{a4}\rangle)  \\
&\xrightarrow{line b}\frac{1}{2}(\frac{1}{2}(|H_{b1}+V_{b1}\rangle\ |H_{b1}-V_{b1}\rangle)-\frac{1}{2}(|H_{b2}+V_{b2}\rangle\ |H_{b2}-V_{b2}\rangle)) \\
&+\frac{1}{2}(|H_{b3}+V_{b3}\rangle\ |H_{b3}-V_{b3}\rangle)-\frac{1}{2}(|H_{b4}+V_{b4}\rangle\ |H_{b4}-V_{b4}\rangle)   \\
&=\frac{1}{\sqrt{8}}(|H_{b1}H_{b1}\rangle-|V_{b1}V_{b1}\rangle-|H_{b2}H_{b2}\rangle+|V_{b2}V_{b2}\rangle+|H_{b3}H_{b3}\rangle-|V_{b3}V_{b3}\rangle-|H_{b4}H_{b4}\rangle+|V_{b4}V_{b4}\rangle)  \\
&\xrightarrow{line c}\frac{1}{\sqrt{8}}(|H_{c1}H_{c1}\rangle-|V_{c2}V_{c2}\rangle-|H_{c4}H_{c4}\rangle+|V_{c3}V_{c3}\rangle+|H_{c5}H_{c5}\rangle-|V_{c6}V_{c6}\rangle-|H_{c7}H_{c7}\rangle+|V_{c8}V_{c8}\rangle) \\
&\xrightarrow{line d}\frac{1}{2}(|H_{d1}V_{d1}\rangle-|H_{d2}V_{d2}\rangle+|H_{d3}V_{d3}\rangle-|H_{d4}V_{d4}\rangle) \\
&\xrightarrow{} C_{D1,D2}+C_{D3,D4}+C_{D5,D6}+C_{D7,D8}.
\end{aligned}
\end{split}
\end{equation}

6. If the input state is $\Psi_{43}=\frac{1}{2}(|03\rangle-|21\rangle+|12\rangle-|30\rangle)$, then we can rewrite the state before each line
\begin{equation}
\begin{split}
\begin{aligned}
&\Psi_{43}=\frac{1}{2}(|03\rangle-|21\rangle+|12\rangle-|30\rangle)  \\
&\xrightarrow{line a}\frac{1}{2}(|H_{a1}V_{a4}\rangle-|H_{a2}V_{a3}\rangle+|H_{a4}V_{a1}\rangle-|V_{a2}H_{a3}\rangle)  \\
&\xrightarrow{line b}\frac{1}{2}(\frac{1}{2}(|H_{b1}+V_{b1}\rangle\ |H_{b2}-V_{b2}\rangle)-\frac{1}{2}(|H_{b2}+V_{b2}\rangle\ |H_{b1}-V_{b1}\rangle) \\
&+\frac{1}{2}(|H_{b3}-V_{b3}\rangle\ |H_{b4}+V_{b4}\rangle)-\frac{1}{2}(|H_{b4}-V_{b4}\rangle\ |H_{b3}+V_{b3}\rangle)   \\
&=\frac{1}{\sqrt{8}}(|H_{b2}V_{b1}\rangle-|H_{b1}V_{b2}\rangle+|H_{b3}H_{b4}\rangle-|H_{b4}V_{b3}\rangle)  \\
&\xrightarrow{line c}\frac{1}{2}(|V_{c2}H_{c4}\rangle-|H_{c1}V_{c3}\rangle+|H_{c5}V_{c8}\rangle-|V_{c6}H_{c7}\rangle) \\
&\xrightarrow{line d}\frac{1}{2}(|V_{d1}H_{D2}\rangle-|H_{d1}V_{d2}\rangle+|H_{d3}V_{d4}\rangle-|H_{d4}V_{d3}\rangle) \\
&\xrightarrow{} C_{D2,D3}+C_{D1,D4}+C_{D5,D8}+C_{D6,D7}.
\end{aligned}
\end{split}
\end{equation}

7. If the input state is $\Psi_{21}=\frac{1}{2}(|01\rangle+|10\rangle+|23\rangle+|32\rangle)$, then we can rewrite the state before each line
\begin{equation}
\begin{split}
\begin{aligned}
&\Psi_{21}=\frac{1}{2}(|01\rangle+|10\rangle+|23\rangle+|32\rangle)  \\
&\xrightarrow{line a}\frac{1}{2}(|H_{a1}V_{a3}\rangle+|V_{a1}H_{a3}\rangle+|H_{a2}V_{a4}\rangle+|V_{a2}H_{a4}\rangle)  \\
&\xrightarrow{line b}\frac{1}{2}(\frac{1}{2}(|H_{b1}+V_{b1}\rangle\ |H_{b1}-V_{b1}\rangle)+\frac{1}{2}(|H_{b2}+V_{b2}\rangle\ |H_{b2}-V_{b2}\rangle) \\
&+\frac{1}{2}(|H_{b3}+V_{b3}\rangle\ |H_{b3}-V_{b3}\rangle)+\frac{1}{2}(|H_{b4}+V_{b4}\rangle\ |H_{b4}-V_{b4}\rangle)   \\
&=\frac{1}{\sqrt{8}}(|H_{b1}H_{b1}\rangle-|V_{b1}V_{b1}\rangle+|H_{b2}H_{b2}\rangle-|V_{b2}V_{b3}\rangle+|H_{b3}H_{b3}\rangle-|V_{b3}V_{b3}\rangle+|H_{b4}H_{b4}\rangle-|V_{b4}V_{b3}\rangle  \\
&\xrightarrow{line c}\frac{1}{\sqrt{8}}(|H_{c1}H_{c1}\rangle+|V_{c2}V_{c2}\rangle-|H_{c4}H_{c4}\rangle-|V_{c3}V_{c2}\rangle+|H_{c5}H_{c5}\rangle+|V_{c6}V_{c6}\rangle-|H_{c7}H_{c7}\rangle-|V_{c8}V_{c8}\rangle) \\
&\xrightarrow{line d}\frac{1}{\sqrt{8}}(|V_{d1}V_{d1}\rangle+|H_{d1}H_{d1}\rangle-|V_{d2}V_{d2}\rangle-|H_{d2}H_{d2}\rangle+|V_{d3}V_{d3}\rangle+|H_{d3}H_{d3}\rangle-|V_{d4}V_{d4}\rangle-|H_{d4}H_{d4}\rangle) \\
&\xrightarrow{} C_{D1,D1}+C_{D2,D2}+C_{D3,D3}+C_{D4,D4}+C_{D5,D5}+C_{D6,D6}+C_{D7,D7}+C_{D8,D8},
\end{aligned}
\end{split}
\end{equation}
here $C_{D1,D1}$ means two photons will arrive at the same detector $D1$. If one has photon-number-resolving detectors, one can determine the state is $\Psi_{21}$ by the measurement result.

Hence, we can divide all the 16 states into 7 classes, such as

1.
\begin{equation}\
\begin{split}
\begin{aligned}
\Psi_{11}=\frac{1}{2}(|00\rangle+|11\rangle+|22\rangle+|33\rangle),  \\
\Psi_{31}=\frac{1}{2}(|02\rangle+|13\rangle+|20\rangle+|31\rangle),  \\
\end{aligned}
\end{split}
\end{equation}
the measured evidence is $C_{D1,D5}+C_{D2,D6}+C_{D3,D7}+C_{D8,D4}$,

2.
\begin{equation}\
\begin{split}
\begin{aligned}
\Psi_{12}=\frac{1}{2}(|00\rangle-|11\rangle+|22\rangle-|33\rangle),  \\
\Psi_{32}=\frac{1}{2}(|02\rangle-|13\rangle+|20\rangle-|31\rangle),  \\
\end{aligned}
\end{split}
\end{equation}
the measured evidence is $C_{D1,D7}+C_{D2,D8}+C_{D5,D3}+C_{D6,D4}$,

3.
\begin{equation}\
\begin{split}
\begin{aligned}
\Psi_{13}=\frac{1}{2}(|00\rangle+|11\rangle-|22\rangle-|33\rangle),  \\
\Psi_{33}=\frac{1}{2}(|02\rangle+|13\rangle-|20\rangle-|31\rangle),  \\
\end{aligned}
\end{split}
\end{equation}
the measured evidence is $C_{D1,D6}+C_{D2,D5}+C_{D8,D4}+C_{D3,D7}$,

4.
\begin{equation}\
\begin{split}
\begin{aligned}
\Psi_{14}=\frac{1}{2}(|00\rangle-|11\rangle-|22\rangle+|33\rangle),  \\
\Psi_{34}=\frac{1}{2}(|02\rangle-|13\rangle-|20\rangle+|31\rangle),  \\
\end{aligned}
\end{split}
\end{equation}
the measured evidence is $C_{D1,D8}+C_{D2,D7}+C_{D4,D5}+C_{D3,D6}$,

5.
\begin{equation}\
\begin{split}
\begin{aligned}
\Psi_{23}=\frac{1}{2}(|01\rangle+|10\rangle-|23\rangle-|32\rangle),  \\
\Psi_{24}=\frac{1}{2}(|01\rangle-|10\rangle-|23\rangle+|32\rangle),  \\
\end{aligned}
\end{split}
\end{equation}
the measured evidence is $C_{D1,D2}+C_{D3,D4}+C_{D5,D6}+C_{D7,D8}$,

6.
\begin{equation}\
\begin{split}
\begin{aligned}
\Psi_{43}=\frac{1}{2}(|03\rangle+|12\rangle-|21\rangle-|30\rangle),  \\
\Psi_{44}=\frac{1}{2}(|03\rangle-|12\rangle-|21\rangle+|30\rangle),  \\
\end{aligned}
\end{split}
\end{equation}
the measured evidence is $C_{D2,D3}+C_{D1,D4}+C_{D5,D8}+C_{D6,D7}$,

7.
\begin{equation}\
\begin{split}
\begin{aligned}
\Psi_{23}=\frac{1}{2}(|01\rangle+|10\rangle-|23\rangle-|32\rangle),  \\
\Psi_{24}=\frac{1}{2}(|01\rangle-|10\rangle-|23\rangle+|32\rangle),  \\
\Psi_{41}=\frac{1}{2}(|03\rangle+|12\rangle+|21\rangle+|30\rangle),  \\
\Psi_{42}=\frac{1}{2}(|03\rangle-|12\rangle+|21\rangle-|30\rangle),  \\
\end{aligned}
\end{split}
\end{equation}
the measured evidence is $C_{D1,D1}+C_{D2,D2}+C_{D3,D3}+C_{D4,D4}+C_{D5,D5}+C_{D6,D6}+C_{D7,D7}+C_{D8,D8}$.

In our experiment, we choose $\Psi_{11}$, $\Psi_{12}$, $\Psi_{13}$, $\Psi_{14}$, $\Psi_{23}$ for quantum superdense coding.

\begin{figure}
\centering
\includegraphics[width=0.6\textwidth]{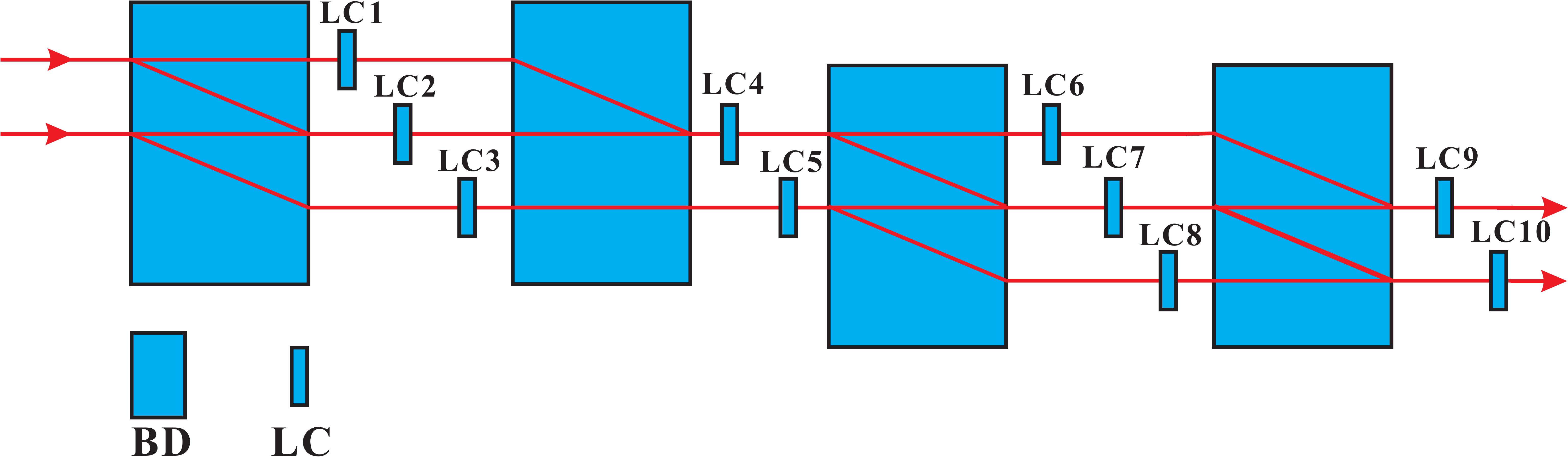}
\caption{{\bf Realization of single photon four-dimensional identical operation and $U$ gate.} The optical axes of liquid crystal variable retarders (LCs) are all set at $45^\circ$. When all the LCs are set to introduce $\pi$ phase, the setup realize identical operation. When LC2, LC4, LC5, LC7, LC9, and LC10 are set to introduce $0$ phase, while the others are set to introduce $\pi$ phase, the setup realize $U$ gate. BD-beam displacer.}
\end{figure}

In fact, one can distinguish six classes of state even without number-resolving detectors. To encode the first five states for SDC, one needs four computer-controlled liquid crystal variable retarders (LCs) to realize the state transformation. In order to encode the sixth state, one needs other ten LCs and four beam displacers as shown in Fig. A2. This part can realize a single photon four-dimensional identical operation or $U$ gate \cite{Babazadeh}. The matrix form of $U$ gate can be written as
\begin{equation}\
\begin{split}
\begin{aligned}
U=\left(
      \begin{array}{cccc}
        0 & 0 & 0 & 1 \\
        0 & 0 & 1 & 0 \\
        0 & 1 & 0 & 0 \\
        1 & 0 & 0 & 0 \\
      \end{array}
    \right).
\end{aligned}
\end{split}
\end{equation}
Together with the first four LCs, Alice can encode six states for SDC. For example, if Alice wants to encode the first five states, she can set the second part to realize an identical operation, if she wants to encode the sixth state $\Psi_{43}$, she needs to transform the initial state $\Psi_{11}$ to $\Psi_{13}$ by the first part, and then realize a $U$ gate by the second part. However, the second part is too complicated and we only use five states in our experiment as a demonstration.
\end{document}